\documentstyle[prd,aps,floats,tabularx,epsfig]{revtex}

\newcommand{\bn}{{\bf n}}

\newcommand{\cd}{\cdot}

\newcommand{\al}{\alpha}

\newcommand{\de}{\delta}
\newcommand{\De}{\Delta}

\newcommand{\ga}{\gamma}
\newcommand{\Ga}{\Gamma}

\newcommand{\la}{\lambda}
\newcommand{\Om}{\Omega}

\newcommand{\si}{\sigma}

\newcommand{\ra}{\rightarrow}

\newcommand{\be}{\begin{equation}}
\newcommand{\ee}{\end{equation}}
\newcommand{\gsim}{\stackrel{>}{\sim}}
\newcommand{\lsim}{\stackrel{<}{\sim}}
\newcommand{\bea}{\begin{eqnarray}}
\newcommand{\eea}{\end{eqnarray}}
\newcommand{\bean}{\begin{eqnarray*}}
\newcommand{\eean}{\end{eqnarray*}}

\newcommand{\bx}{{\bf x}}
\newcommand{\Gauss}{{\mbox{Gauss}}}

\begin{document}
\draft
\preprint{\ 
\begin{tabular}{rr}
& 
\end{tabular}
} 
\twocolumn[\hsize\textwidth\columnwidth\hsize\csname@twocolumnfalse\endcsname 
\title{Tensor Microwave Anisotropies from a Stochastic Magnetic Field 
}
\author{ R. Durrer$^1$, P.G. Ferreira$^{1,2,3}$ 
and T. Kahniashvili$^4$ }
\address{ $^1$D\'epartement de Physique Th\'eorique, Universit\'e de Gen\`eve,
24 quai Ernest Ansermet, CH-1211 Gen\`eve 4, Switzerland\\
$^2$CERN Theory Division, CH-1211, Geneve 23, Switzerland\\
$^3$CENTRA, Instituto Superior T\'ecnico, Av. Rovisco Pais, 1, 1096 Lisboa Codex, Portugal\\
$^4$Department of Astrophysics, Abastumani 
Astrophysical Observatory,
Kazbegi Ave. 2a,  380060 Tbilisi, Georgia}
\maketitle

\begin{abstract}
We derive an expression for the angular power spectrum of cosmic
microwave background anisotropies due to gravity waves generated
 by a stochastic magnetic 
field and compare the result with current observations; we take into
account the non-linear nature of the stress energy tensor of the
magnetic field.
 For almost scale invariant spectra, the 
amplitude of the magnetic field at galactic scales is constrained to
be of order $10^{-9}$ Gauss. If we assume that the magnetic field is
damped below the Alfv\'en damping scale, we find that its amplitude at
 $0.1$h$^{-1}$Mpc, $B_\lambda$, is constrained to be
$B_\lambda<7.9\times10^{-6} e^{3n}$ Gauss, for $n<-3/2$, and 
$B_\lambda<9.5\times10^{-8} e^{0.37n}$ Gauss, for $n>-3/2$,
where $n$ is the spectral
index of the magnetic field and $H_0=100h$km s$^{-1}$Mpc$^{-1}$ is the
Hubble constant today.
\end{abstract}
\date{\today}
] 
\renewcommand{\thefootnote}{\arabic{footnote}} \setcounter{footnote}{0}
\section{Introduction}
The past few years have seen a tremendous surge of interest in the
origin and evolution of galactic magnetic fields \cite{kronberg}. A number of 
mechanisms have been proposed for the origin
of the seed fields, ranging from inflationary mechanisms 
\cite{inflation}, cosmological
phase transitions \cite{phase} to
 astrophysical processes \cite{astro}. Much progress
has been made in trying to disentangle the various non-linear processes
which may be responsible for the growth of such a seed field
in the very early universe in particular the interplay between the
magnetic field and the primordial plasma \cite{mhd,sblong} and the
importance of turbulence\cite{invcascade}. 

Given a small seed field at late times, two different
mechanisms can cause its amplification to magnetic fields of
order $10^{-6}$ Gauss observed in  galaxies: adiabatic compression of
magnetic flux lines can amplify a seed field of order $10^{-9}$ Gauss
to the present, observable values; the far more efficient (and controversial)
galactic dynamo mechanism may be able to amplify seed fields as 
small as 10$^{-20}$Gauss \cite{astro} or even 10$^{-30}$Gauss
in universe with low mass density \cite{davis}.
Clearly, to make some progress in identifying which one of these
mechanisms is responsible for galactic magnetic fields, one would
like to find a constraint for the seed field before it has been
processed by local, galactic dynamics. 

The obvious observable for such a constraint is the cosmic microwave
background (CMB). It is interesting to note that a field strength of 
$10^{-8}$Gauss provides an energy density of $\Om_B = B^2/(8\pi\rho_c) \sim
10^{-5}\Om_{\ga}$, where $\Om_{\ga}$ is the density parameter in
photons. We naively expect a magnetic field of this amplitude to 
induce perturbations in the CMB on the order of $10^{-5}$, which are 
just on the level of the observed CMB anisotropies. It is thus 
justified to wonder to what extent the isotropy of the CMB may 
constrain primordial magnetic fields.
Our order of magnitude estimate makes clear that we shall never be
able to constrain tiny seed fields on the order of $10^{-13}$Gauss or
less in this way, but primordial fields of $10^{-9}$Gauss may have
left their traces in the CMB. 

A number of methods have been proposed
in the past few years for measuring a cosmological magnetic field
using the CMB: the effect on the acoustic peaks  \cite{jenni}, Faraday
rotation on small  \cite{kosowsky} and large   \cite{evan} scales and vorticity
\cite{sb,RTA}  can all lead to observable 
anisotropies in the CMB if the primordial magnetic field strength is of
the order of $10^{-9}$ to $10^{-8}$ Gauss.
The most stringent bound from the CMB presented thus far was for the
case of a homogeneous magnetic field\cite{BFS}; the
authors  use the COBE data to find the constraint 
$B_0<6.8\times10^{-9}(\Omega_0 h^2)^{1/2}$Gauss where the Hubble
constant is $H_0=100h$km s$^{-1}$Mpc$^{-1}$ and $\Omega_0$ is
the energy density in units of the critical value.
Although there is no fundamental reason to discard the possibility
of a homogeneous magnetic field, all physical mechanisms proposed
to date lead to the presence of stochastic magnetic fields with 
no homogeneous term;
in this paper we consider such fields. For these types of
configuration one is allowed to have fluctuations on a wide range
of scales and the magnetic field will serve as a non-linear driving force
to the metric fluctuations; in the parlance of cosmological perturbation
theory, the magnetic field evolves as a {\it stiff source}, without 
being affected by the fluid perturbations (back reaction)\cite{fc}
which may be induced.

Stochastic magnetic fields have also been considered in~\cite{sb},
where the CMB anisotropies due to the induced fluid vorticity has been
analyzed. Here we determine gravitational effects of the magnetic
field. For simplicity, and to allow for a purely analytical analysis,
we constrain ourselves to tensor
perturbations. Similar contributions are also expected from vector and
scalar perturbations which then would add to the final result. In this
sense the anisotropies computed here are a strict lower bound
(underestimating the true effect probably by about a factor of three).

The main result of this work is that one can obtain
reasonably tight constrains for scale invariant magnetic fields;
For causally generated magnetic fields the constraints are 
weaker and are strongly dependent on the evolution
of the magnetic field in the radiation era on small scales.

For simplicity we concentrate on the case $\Omega_0=1$. Througout, we
use conformal time which we denote by $\eta$. 
Greek indices run from $0$ to $3$, Latin ones from $1$ to $3$.
We denote spatial (3d) vectors with bold face symbols. The value of 
the scale factor today is $a(\eta_0)=1$.

\section{ The stress tensor of the magnetic field}
During the evolution of the universe, the 
conductivity of the inter galactic medium is effectively infinite.
 In this regime we
can decouple the time evolution from the spatial structure: 
{\bf B} scales like ${\bf B}(\eta,{\bf x})={\bf B}_0({\bf x})/a^2$ on
sufficiently large scales. On smaller scales the interaction of the
magnetic field with the cosmic plasma becomes important leading mainly
to two effects: on intermediate scales, the oscillates like
$\cos(v_Ak\eta)$, where $v_A = B^2/(4\pi(\rho+p))^{1/2}$ is the
Alfv\'en velocity and on small scales, the field is exponentially
damped due to shear viscosity~\cite{sblong}.

We will model ${\bf B}_0({\bf x})$ as a statistically homogeneous
and isotropic random field. The transversal nature of ${\bf B}$
leads us to
\be
  \langle B_i({\bf k})B^*_j({\bf q})\rangle = 
        \de^3({\bf k-q})(\de_{ij}-\hat{k}_i\hat{k}_j)B^2(k)~. \label{powB}
\ee
where we use the Fourier transform conventions
\bean
\ B_j({\bf k}) &=&\int d^3x 
\exp(i{\bf x\cd k})B_{0j}({\bf x}) ~,\\
  B_{0j}({\bf x}) &= & {1\over (2\pi)^3}\int d^3k \exp(-i{\bf x\cd k})
         B_{j}({\bf k}) ~.
\eean
The Alfv\'en oscillations modulate the initial power spectrum by a
factor
\[ B^2(k)\ra B^2(k)\cos^2(v_Ak\eta) ~.\]
This can be approximated by a reduction of a factor $2$ in the power
spectrum on scales with $v_Ak\eta \gsim 1$. But as we shall see, our
most stringent constraints will come either from very small scales
where the spectrum is exponentially damped or from much larger scales where
oscillations can be ignored. We will incorporate the exponential damping by
a cutoff in the power spectrum at the damping scale.

Let us investigate the consequence of causality for the spectrum
$B^2(k)$.  If $\bf B$ is
generated by some causal mechanism, it is uncorrelated on super
horizon scales,
\be
 \langle B_i({\bf x},\eta)B_j({\bf x}',\eta)\rangle=0  ~~~{\rm for }~~~ 
|{\bf x}-{\bf x}'| > 2\eta ~.  \label{causal}
\ee
Here it is important that the universe is in a stage of standard
Friedmann  expansion, so that the causal horizon size is about $\eta$.
During an inflationay phase the causal horizon diverges and our
subsequent argument does not apply. In this somewhat misleading sense,
one calles inflationary perturbations 'a-causal'.

According to Eq.~(\ref{causal}),
$ \langle B_i({\bf x},\eta)B_j({\bf x}',\eta)\rangle$ is a
 function with compact support and hence its Fourier
 transform is analytic. The function
\be
  \langle B_i({\bf k})B^*_j({\bf k})\rangle \equiv 
 (\de_{ij}-\hat{k}_i\hat{k}_j)B^2(k)
\ee
is analytic in {\bf k}. If we in addition assume that $B^2(k)$ can be
approximated by a simple power law, we must conclude that $B^2(k)
 \propto k^n$, where $n\ge 2$ is a even integer. (A white noise
spectrum, $n=0$ does not work because of the transversality condition
which has led to the non-analytic pre-factor $\de_{ij}-\hat{k}_i\hat{k}_j$.)
By causality, there can be no deviations of this law on scales larger
than the horizon size at formation, $\eta_{in}$.

We assume that the probability distribution function of 
${\bf B}_0$ is Gaussian; although this is not the most general random 
field, it greatly simplifies calculations and  gives us a good
idea of what to expect in a more general case. 

The anisotropic stresses induced  are given by the
convolution of the  magnetic field,
\bea
\tau_{ij}^{(B)}({\bf k}) &=& {1\over 4\pi}\int d^3qB_i({\bf q})B_j^*({\bf
        k-q}) \nonumber \\
 && - {1\over 2}B_l({\bf q})B_l^*({\bf k-q})\de_{ij}~.
\label{taub}
\eea
With the use of the projection operator, 
$P_{ij} = \de_{ij}-\hat{k}_i\hat{k}_j$ we can extract the tensor component 
of Eq.~(\ref{taub}), 
\be
\Pi_{ij}^{(B)}= (P_i^aP_j^b-(1/2)P_{ij}P^{ab})\tau_{ab}~,
\label{pib}
\ee
tracelessness, orthogonality and symmetry force the correlation
function to be of the form 
\bea
\langle \Pi^{(B)}_{ij}({\bf k},t)\Pi^{(B)*}_{lm}({\bf
        k}',t) \rangle &=&|\Pi_B(k,t)|^2{\cal M}_{ijlm}\de({\bf
k-k'})\nonumber \\
\langle \Pi^{(B)}_{ij}({\bf k},t) \Pi^{(B)*}_{ij}({\bf k'},t)\rangle
&=& 4|\Pi_B(k,t)|^2\de({\bf k-k'}), 
 \label{powPi}
\eea
were we make use of the tensor basis, ${\cal M}$: The correlator on an
 isotropic tensor component has always the following tensorial structure,
\bea
{\cal M}_{ijlm}&=&\de_{il}\de_{jm}+\de_{im}\de_{jl}   -\de_{ij}\de_{lm} + 
        k^{-2}(\de_{ij}k_lk_m + \nonumber \\ 
&&      \de_{lm}k_ik_j -\de_{il}k_jk_m - \de_{im}k_lk_j -\de_{jl}k_ik_m
        \nonumber \\
&&      -\de_{jm}k_lk_i) + k^{-4}k_ik_jk_lk_m~.
\eea
We now determine the function $|\Pi_B(k,t)|^2$ in terms of the
magnetic field. Using Wick's theorem we have
\bea 
\lefteqn{\langle B_i({\bf k})B^*_j({\bf q}) B_n({\bf s})B^*_m({\bf p})\rangle 
  = } \nonumber \\  &&
\langle B_i({\bf k})B^*_j({\bf q})\rangle
\langle B_n({\bf s})B^*_m({\bf p})\rangle + \nonumber \\ && 
\langle B_i({\bf k})B^*_n({\bf s})
\rangle\langle B_j({\bf q})B^*_m({\bf p})\rangle + \nonumber \\ && 
\langle B_i({\bf k})B^*_m({\bf p})\rangle
\langle B_n({\bf s})B^*_j({\bf q})\rangle ~. \label{corB2}
\eea
 The problem reduces itself to
calculating self convolutions of the magnetic field. The power spectrum of
Eq.~(\ref{taub}) is
\bea
\lefteqn{\langle \tau^{B}_{ij}({\bf k},\eta)\tau^{B*}_{lm}({\bf
k'},\eta)\rangle =} \nonumber\\ 
 &&{1\over (8\pi)^2}\int d^3\! q\! \int d^3 p \langle B_{i}({\bf q})
B_{j}({\bf (k-q)}) 
B_{l}({\bf p})B_{m}({\bf (p-k')}) \rangle =     
\nonumber \\ && 
\delta({\bf k}-{\bf k'}) \int d^3 q~B^2(q) B^2(|{\bf k}-{\bf q}|) \times
\nonumber \\ && 
[ (\delta_{il}- {\hat q}_i {\hat q}_l) 
(\delta_{jm}- {\widehat {({\bf k}-{\bf q})}}_j {\widehat 
{({\bf k}- {\bf q})}}_m)+
\nonumber \\ &&  
(\delta_{im}- {\hat q}_i{\hat q}_m)
(\delta_{jl}- {\widehat{({\bf k}-{\bf q})}}_j 
        { \widehat{({\bf k}-{\bf q})}}_l) ]
~.  \label{ijlm}
\eea
Using Eqs. (\ref{corB2},\ref{pib}) and (\ref{powPi}), this leads to
$|\Pi_B|^2  =  f(k)^2/a^8$,   where
\be 
 f(k)^2 ={1\over (8\pi)^2} \int d^3 q B^2(q) B^2(|{\bf k} - {\bf q}|)
(1+2\gamma^2+\gamma^2 \beta^2) ~, \label{ff}
\ee
with $\gamma={\bf \hat k}\cdot{\bf \hat q}$ 
and  $\beta = {\bf \hat k}\cdot { \widehat{\bf k-q}}$.

It remains to define $B(k)$ from Eq.~(\ref{powB}). We shall parameterize
it in terms of an amplitude and a scale dependence through
\be
B^2(k)= \left\{\begin{array}{ll}
        \frac{(2\pi)^5}{4}
{\lambda^{n+3}\over \Gamma[\frac{n+3}{2}]}B_{\lambda}^2k^n &
        \mbox{ for } k<k_c\\
        0 &     \mbox{ otherwise} \end{array} \right.
\label{magfield}\ee
The normalization is such that $\langle B^i_0({\bf x})B^i_0({\bf x})
\rangle|_\lambda=B^2_\lambda$ where the quantity in brackets represents
the averaged magnetic field smoothed over a comoving length scale $\lambda$.
Note that we have assumed that the cutoff scale today is smaller than
$\lambda$.

We require $n>-3$ so as  not to over-produce long
range coherent fields; we shall see that for $n=-3$ we obtain a
scale invariant spectrum of CMB anisotropies. 

We have included a short wavelength cutoff to take into account the
exponential damping due to shear viscosity in the cosmic plasma
\cite{sblong}.
The mean energy density due to such a magnetic field, which is 
an  appropriately weighted integral of Eq.~(\ref{magfield}),
will be strongly dependent on the cut off when $n>-3$. 

Using Eqs.~(\ref{magfield}) and (\ref{ff}) we can calculate $f$. The
integral cannot be computed analytically, but the following result is
a good approximation for all wave numbers $k$
\be
  f^2(k) \simeq {(2\pi)^9\over 16} \frac{\lambda^{2n+6}B_\lambda^4} 
{\Gamma^2[\frac{n+3}{2}](2n+3)}\left(
        k_c^{2n+3} +\frac{n}{n+3}k^{2n+3}\right) ~.
\label{faprox}
\ee
This result seems to have a singularity at $n=-3/2$ which is however
removable. The first term dominates if $n>-3/2$ and while second term
dominates if the opposite inequality is satisfied. For $n>-3/2$, the
gravity wave source is therefore white noise and its amplitude is
determined by the upper cutoff, $k_c$.  Note that if $n>-3/2$, the
spectrum of the energy momentum tensor becomes white noise,
independent of $n$. Only the amplitude which is proportional to 
$(\la k_c)^{2n}$ depends on the spectral index. This is due to the
fact that the integral (\ref{ff}) is dominated by the contributions at
very small scales, $k_c\gg k$. The induced $C_\ell$ spectrum from
gravity wave  will therefore be independent of $n$ for $n>-3/2$, and
obey the well known behavior $C_\ell \propto \ell$ of a white noise
source.

To simplify, we
just consider the dominant term and, in order not to
artificially produce a singularity at $n=-3/2$, we drop the factor $1/(2n+3)$.
Given the intent of this paper (to constrain the amplitude of the
magnetic field) we will include a factor of $10^{-1}$ in our final
result, guaranteeing that we are not overestimating CMB anisotropies.
The singularity at $n=-3$ is real. It is the usual logarithmic
singularity of the scale invariant spectrum.

\section{ The CMB anisotropies}
Armed with the structure and evolution of the stochastic magnetic field
we can now proceed to calculate its effects on tensor CMB anisotropies.
The metric element of the perturbed Friedman universe is given  by
\[ds^{2}=a^{2}(\eta)(-d\eta^2+(\delta_{ij}+2h_{ij})
dx^idx^j)~, \]
where $h^i_i=0$ and $h^j_{i}k^{i}=0$ for tensor perturbations.
The magnetic field will source the evolution equation for $h_{ij}$
through
\be
{\ddot{h}}_{ij}+2{\dot a \over a}\dot{h}_{ij}+
 k^2 h_{ij}=8\pi G \Pi_{ij}^{(B)} ~.
\label{hprime}
\ee
Such a gravity wave induces temperature fluctuations in the CMB due to
the fact that the photons move along the perturbed geodesics\cite{fc}
\be
{\De T\over T}(\eta_0,\bx,\bn) =
     \int_{\eta_{*}}^{\eta_0}\dot{h}_{ij}(\bx(\eta),\eta)n^in^j
d\eta~. \label{dT}
\ee
Here $\eta_*$ denotes the (conformal) time of decoupling of matter and
radiation due to recombination.
We want to compute the angular power spectrum of ${\De T\over T}$, the
$C_\ell$, defined by 
\[ \langle{\De T\over T}(\bn){\De T\over T}(\bn') 
        \rangle_{\bn\cd\bn'=\mu}={1\over
        4\pi}\sum_l(2\ell+1)C_{\ell}P_\ell(\mu)~. \]
The $C_\ell$s are solely determined by the power spectrum of
 metric fluctuations. 
Defining 
\[ \langle \dot{h}^{(T)}_{ij}({\bf k'},\eta)\dot{h}^{(T)*}_{lm}({\bf
        k},\eta)\rangle = |\dot{H}(k,\eta)|^2{\cal M}_{ijlm}\de({\bf k-k'})\]
one can derive a closed form expression for $C_\ell$ (see \cite{AbottWise}):
\bea
C_\ell &=&{1\over 4\pi^4}\int dkk^2|I(\ell,k)|^2
\ell(\ell-1)(\ell+1)(\ell+2) ~,
\label{Clt}\\
I&=&\int_{\eta_*}^{\eta_0}d\eta\dot{H}(\eta,k)
        \left({j_{\ell}(k(\eta_0-\eta)) \over (k(\eta_0-\eta))^2} \right )
\label{II}
\eea
where $j_{\ell}$ denotes the spherical Bessel function of order $\ell$. 
We solve equation (\ref{hprime}) using the Wronskian method; in terms of
 the dimensionless variable $x=k\eta$. The homogeneous
solutions are the  spherical
Bessel functions $j_0~,y_0$ in the radiation  dominated era,
and $j_1/x~,y_1/x$ in the matter dominated era respectively. 
We assume that the magnetic fields were created in the radiation
dominated epoch, at redshift $z_{in}$. We then match 
the general inhomogeneous solutions of
Eq.~(\ref{hprime})  at the time of equal matter and 
radiation, $\eta_{eq}$. Due to the rapid falloff of the source term in
the matter dominated era, the
perturbations created after $\eta_{eq}$ are sub-dominant, and we find
for the dominant contribution at $\eta>\eta_{eq}$
\bea
 \dot{H}(k,t) &\simeq& 4\pi G\eta_0^2 z_{eq}\ln
\left(\frac{z_{in}}{z_{eq}}\right)kf(k)
        {j_2(k\eta)\over k\eta}
 \label{Hsol'} ~.
\eea
Inserting this result in  Eq.~(\ref{II}), we obtain
\be
I = 4\pi G\eta_0^2z_{eq}\ln(z_{in}/z_{eq}) f(k)
\int_{x_{*}}^{x_0}dx {j_2(x) \over x} 
{j_{\ell}(x_0-x) \over (x_0-x)^2} ~,
\label{II1}
\ee
where $x=k\eta$, $x_* = k\eta_*$ and $x_0=k\eta_0$.
For wave numbers which are super-horizon at decoupling, $x_* < \pi$, 
the lower boundary in Eq.~(\ref{II1}) 
 can be set to $0$. The remaining integral cannot be expressed in closed form,
but is  well approximated by \cite{AS}:
\bea
\lefteqn{\int_{0}^{x_0}dx {j_2(x) \over x} 
        {j_{\ell}(x_0-x) \over (x_0-x)^2}  =} \nonumber\\
&& { \pi \over 2}\int_{0}^{x_0}dx {J_{5/2}(x) \over x^{3/2}} 
{J_{\ell+1/2}(x_0-x) \over (x_0-x)^{5/2}} 
\simeq \nonumber \\ &&
        {0.7 \pi\over 2}\int_{0}^{x_0}dx {J_{5/2}(x) \over x} 
        {J_{\ell+1/2}(x_0-x) \over (x_0-x)^3} \nonumber \\ &&
 = {7\pi\over 25}{\sqrt{\ell}\over x_0^3} J_{\ell+3}(x_0))~.
\eea
The third integral above   can be expressed in closed form (\cite{GR}, 
number 6.581.2), and is reasonably well approximated by the last
expression,  we have checked the approximation numerically
 for  $l\le 200$ and varying $x_0$.

We can now do the integrals in  Eq.~(\ref{Clt}) analytically to obtain
\be
\hspace{-0.7in}\ell^2C_\ell \simeq A \left(
        \frac{\lambda}{\eta_0}\right )^{2n+6}
   {2\over 3\pi}(k_c\eta_0)^{2n+3}\ell^3
\label{clplus}
\ee
for $n>-3/2$, and
\be
\ell^2C_\ell \simeq A \left (\frac{\lambda}{\eta_0}\right )^{2n+6}
 {-n\over (n+3)}{\Ga[1-2n]\over\Ga^2[1-n]2^{(1-2n)}} \ell^{6+2n}
\label{clminus}
\ee 
for $-3<n<-3/2$, where 
\bea
A &=& 5\times 10^{-4} (2\pi)^{9} z_{eq}^2\ln^2
  \left({z_{in}\over z_{eq}}\right){B_\lambda^4G^2\eta_0^4\over 
\Gamma^2[\frac{n+3}{2}]}\nonumber \\
&=&  3\times 10^{-8} \left( {B_\lambda\over
10^{-9}\Gauss}\right)^4\ln^2(z_{in}/z_{eq})
\frac{1}{\Gamma^2[\frac{n+3}{2}]}~.
\eea


\section{Results}
Eqs.~(\ref{clplus}) and (\ref{clminus}) are our main result. 
They allow us to limit a possible 
primordial magnetic field by requiring it not to over produce fluctuations 
in the CMB. Since the
fluctuations induced grow with $\ell$ for all values of the spectral
index $-3< n$, we obtain the best limits for large values of
$\ell$. We shall be conservative and assume an upper bound of
$\ell^2C_\ell \mid_{\ell=50}<8.5\times 10^{-9}$ \cite{cmbconst}.
Given that we are interested in galactic and cluster scales
we  fix $\lambda=0.1$ h$^{-1}$Mpc for the remainder of this paper.
In Fig.~1 we show the limit on a stochastic magnetic field as a function
of the spectral index $n$, using the damping scale given below as cutoff.

We  now focus on a few particular cases of interest and in
doing so we will derive an analytic expression which approximates
the upper bound of $B_\lambda$ over the whole range of $n$.

{\it Scale invariant magnetic field}: From Eq (\ref{clminus}) we see that
the result is independent of the cutoff. In the limit where $n\rightarrow-3$
we find that
\begin{eqnarray}
B_{\lambda}\lsim 10^{-9}{\rm Gauss}
\end{eqnarray}
i.e. of the same level as other 
constraints~\cite{jenni,kosowsky,evan,sb,RTA,BFS}.

{\it Causal magnetic field}: For this scenario we have, as explained
above, $n\ge2$; we shall consider the case of $n=2$. For instructive
purposes let us first consider a $k_c$ which is independent of
the magnetic field. The constraint
is then
\begin{eqnarray}
B_{\lambda}\lsim \ln^{-\frac{1}{2}}\left (\frac{z_{in}}{z_{eq}}\right)
(k_c\eta_0)^{-\frac{7}{4}}{\rm Gauss}~. \label{causal1}
\end{eqnarray}
The cutoff $k_c$ will 
depend on the plasma properties and evolution;
even though the conductivity
$\si$ of the cosmic plasma is very large, it is nevertheless
finite. One actually finds\cite{ae} that $\si = \al T$,
where the parameter $1< \al <7$ is slowly temperature dependent. By
Ohm's law, magnetic fields on small enough scales are exponentially
suppressed, $ B\propto \exp\left({-a k^2\eta/ 4\pi\si}\right)$,
leading to a damping scale, $k_d(\eta)= (4\pi\si a/\eta)^{1/2}
=(\eta \times2\times 10^{-3} {\rm cm})^{-1/2}$.
This  scale is smaller than the comoving horizon scale for all
temperatures below the Planck scale. On scales smaller than
$1/k_d(\eta_{eq})$, the induced gravity waves have damped
away even before matter and radiation equality.
Since the sourcing of gravity waves after equality is negligible, the
damping scale relevant in our problem is $k_c=k_d(\eta_{eq})$,
\be
 k_c\simeq 2\times 10^{13}h^2{\rm Mpc}^{-1}~. \label{declength}
\ee
If we  insert this this damping scale in Eq.~(\ref{causal1}), we obtain
$B_{\lambda}\lsim 10^{-29}{\rm Gauss}$. 

A more realistic scenario is to assume that
the magnetic field will be damped by electron viscosity. 
To proceed with the analysis we shall split the stochastic magnetic field 
into a high-frequency component and a low-frequency component;
the scale which separates the two is the Alfven scale at
equality, $\lambda_A=V_A\eta_{eq}$ where $V_A$ is the Alfven
velocity, $V_A^2 =\langle B^2\rangle/(4\pi(\rho+p))$.
 From equation 4 of \cite{sb} we see that the inhomogeneous magnetic field
will obey a damped harmonic oscillator equation, with a 
time dependent damping coefficient, $D=0.2k^2l_\gamma(1+z)$
($l_\gamma$ is the physical photon mean free path)
and frequency $\omega(\eta)=V_A^2k^2-({\dot D}/2)-(D/2)^2$.
Within this setting we can estimate the damping scale of
the magnetic field in the oscillatory regime 
of this system; the amplitude
of the effective homogeneous magnetic field, $B_A$,
 which is responsible for
the Alfv\'en waves is related to $B_\lambda$ through
\[ B_A^2 \simeq B_\la^2\left({\la 10^{-9}\mbox{Gauss}\over 3.8\times
10^{-4}B_A\eta_{eq}}\right)^{n+3} \]
which leads to 
\be
B_A = \left (\frac{B_\lambda}{10^{-9}{\rm Gauss}}\right)^\frac{2}{n+5}
(13h^{-1})^\frac{n+3}{n+5}10^{-9}{\rm Gauss}
 \label{decalfven}
\ee
We shall define the damping scale to be the scale at which
one e-fold of damping has occured by equality. From
$\int_0^{\eta_{eq}} \frac{D}{2}d\eta=1$
one finds 
\be
k_c=4.5 {\rm Mpc}^{-1}\label{dampscale}
\ee
For this estimate to be valid, the system must be in
the damped oscillatory regime (as opposed to overdamping regime), i.e.
 $\omega(\eta_{eq})>0$; this condition is satisfied
if $B_A>5.5h^{-2}\times 10^{-9}$ Gauss. We find that indeed this 
is the case in the range of interest.

 Combining Equations (\ref{decalfven}),(\ref{clplus}),
(\ref{clminus}) and (\ref{dampscale}) and assuming a formation
redshift of $z_{in}=10^{15}$ (although the final result
is very weakly dependent $z_{in}$) we find that an
an approximation to the bound is
\bea
B_\lambda&<&7.9\times10^{-6} e^{2.99n} {\rm Gauss},\ \ \ \ \mbox{for $n<-3/2$}, 
\nonumber \\ 
B_\lambda&<&9.5\times10^{-8} e^{0.37n} {\rm Gauss},\ \ \ \ \mbox{for $n>-3/2$}
,\label{fit}
\eea
The upper bounds corresponding to Eq. \ref{fit} represent a reasonable
fit to Figure 1. As one can see, the constraint on
a causal magnetic field is well above $10^{-9}$ Gauss.

Throughout this derivation we have assumed that we can estimate
the damping scale of the magnetic field by looking solely at the
Alfven modes. A linear analysis of the remaining degrees of
freedom also indicate that the magnetic field will be damped
at the same scale as in Equation (\ref{dampscale}). It is possible
that non-linear effects may prevent the tangled magnetic field
from damping at this scale but an accurate quantitative analysis 
is still lacking.

\begin{figure}[ht]
\centerline{\epsfxsize=3in  \epsfbox{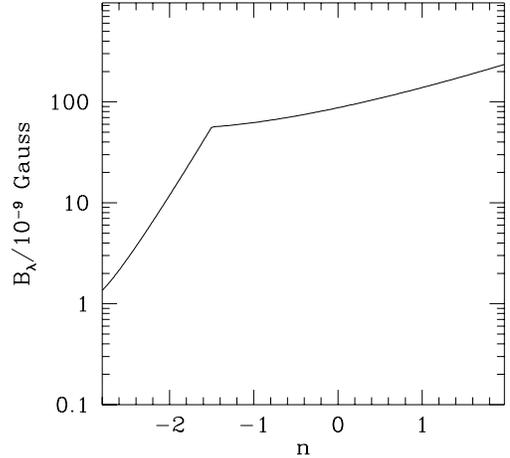}}
\caption{The upper bound $B_\lambda$ as a function of spectral index,
$n$. We assume $z_{in}/z_{eq}=10^{9}$ and $\lambda=0.1$h$^{-1}$Mpc}.
\end{figure}

{\it Inflationary magnetic fields}: Broken conformal invariance
allied with the an inflationary period will create large scale magnetic
fields. Bertolami and Mota \cite{inflation} estimate the spectral index
in such a mechanism to lie in the range around 0. 
We are then clearly in 
the regime where the cutoff is important. Using the Alfv\'en damping scale
at equality and assuming magnetic field generation
at $10^{10}$GeV, we find $B_{\lambda}\lsim (10^{-7}-10^{-8}){\rm Gauss}$
for $n$ varying from $-0.5$ to $0.5$.
A similar result can be obtained for the model of Gasperini {\it et al}
\cite{inflation}.

\section{Discussion}

Our calculation differs from most of the recent work on the impact of 
primordial magnetic fields on structure formation: In estimating
the CMB anisotropies we do not split the magnetic
field into a 'large' homogenous mode and a 'small' fluctuation.
 The magnetic field then affects metric perturbations quadratically.
This has two effects. Firstly it allows us to consider the magnetic field
as a stiff source, and discard (within the MHD approximation) the backreaction
of the perturbations in the cosmological fluid. Indeed if we {\it
were} to consider backreaction 
then we would know a priori that we would be generating 
unacceptable perturbations in the cosmological fluid. 
Another way of phrasing this is that the magnetic field itself is
$1/2$ order perturbation theory, while its energy momentum tensor and
consequently the induced metric perturbations are first order
perturbations. The MHD backreactions on $\bf B$ would be $3/2$ order
and may thus be neglected in linear perturbation theory. We point
out, however that, to obtain an estimate of the damping scale due
to the viscosity in the MHD we had to consider a split between
long wave length  and short wavelength fluctuations in $\bf B$.

Secondly, the stress energy tensor being quadratic in the magnetic
field,  leads to a 'sweeping' of modes: large wavelength
modes in $T_{\mu\nu}$ will in general be affected by all scales of 
the spectrum of $B$ \cite{sb}. As we have seen in the causal case, the
small wavelength behaviour of the magnetic field totally dominates the large
wavelength pertubations.
In \cite{mhd} the magnetic field is modeled as 
${\bf B}={\bf {\bar B}}+{\bf B}^{(1)}({\bf x})$ where
${\bf {\bar B}}$ is a homogeneous term; the stress energy tensor is then
given by  terms of the form ${\bar B}_{i}B_j^{(1)}$, which are linear in the
stochastic component.

A few comments are in order with regards to our result.
Note that we are considering a specific class of models, where
the magnetic field seed is created at some well defined moment in the
early universe and then evolves according to the MHD equations. If
the magnetic field is being constantly sourced throughout the radiation
era, then our calculation is not valid. An example of such a scenario
was proposed by Vachaspati \cite{phase} where magentic fields are
sourced by vortical imprints from an evolving network of cosmic strings;
although the scaling behaviour of source may lead to $B\propto a^{-2}$,
the effective damping scale will be of order the horizon much larger
than the Alfv\'en damping scale.
Another possibility has been put forward in \cite{invcascade},
where the onset of turbulence induces an amplification of power on
large scales but a supression of power on small scales. 
This would further increase $k_c$ but the results are still too qualitative to
be properly included in an analysis such as ours.

{\it Acknowledgments}: We thank John Barrow, Orfeu Bertolami, Kari Enqvist,
Karsten Jedamzik, Jo\~ao Magueijo, Evan Scannapiecco and Misha Shaposhnikov for
useful discussions. We thank one of the referees for pointing
out the importance of Equation (\ref{decalfven}) and another for
alerting us to the importance of reference \cite{sb}.
RD acknowledges the hospitality of the CfPA at
U.C. Berkeley, where this work was initiated. TK acknowledges
the hospitality of Geneva University.
\tighten

\end{document}